# Topological Regularization for Force Prediction in Active Particle Suspension with EGNN and Persistent Homology

Sadra Saremi*, Amirhossein Ahmadkhan Kordbacheh*
Department of Physics, Iran University of Science and Technology, Tehran 16846-13114, Iran.

## Abstract

Capturing the dynamics of active particles, i.e., small self-propelled agents that both deform and are deformed by a fluid in which they move is a formidable problem as it requires coupling fine scale hydrodynamics with large scale collective effects. So we present a multi-scale framework that combines the three learning-driven tools to learn in concert within one pipeline. We use high-resolution Lattice Boltzmann snapshots of fluid velocity and particle stresses in a periodic box as input to the learning pipeline. the second step takes the morphology and positions orientations of particles to predict pairwise interaction forces between them with a E(2)-equivariant graph neural network that necessarily respect flat symmetries. Then, a physics-informed neural network further updates these local estimates by summing over them with a stress data using Fourier feature mappings and residual blocks that is additionally regularized with a topological term (introduced by persistent homology) to penalize unrealistically tangled or spurious connections. In concert, these stages deliver an holistic highly-data driven full force network prediction empathizing on the physical underpinnings together with emerging multi-scale structure typical for active matter.

## Introduction

Self propelled active particle suspensions — which encompasses, e.g., bacteria, algae or synthetic colloidal rotors that eat power rather than momentum [1][2][3]—display a wide variety of non-equilibrium phenomena that emerge from a coupling among an individual's propulsion, inter-particle interactions and hydrodynamics. Although continuum models built on both kinetic theory and coarse grained hydrodynamics have been used to describe large scale instabilities and pattern formation in bacterial turbulence [4][5][6], they typically have insufficient resolution to resolve pairwise forces [24], as well as the emergent network topology on the particle scale [7][8][9]. They are often used for coupling between fluid and particle computations in high Reynolds number flows, which requires unstructured meshes when doing so [7][8][9]; however extracting the forces directly from these simulations as system size grows quickly becomes computationally infeasible. Recent advances in machine learning may allow us to infer effective interaction rules for active matter.

Other geometric deep learning frameworks, such as graph neural networks (GNNs) [13][12][34], have been developed in order to learn relational inductive biases from material physical systems, and group equivariant architectures enforce the learned predictions to be consistent w.r.t. Euclidean symmetries [10][11][33] in parallel to Physics-informed neural networks (PINNs) [14–17] incorporate governing equations into the loss function to enforce that the learned solution satisfies the relevant physical consistency terms (for example, Navier–Stokes residuals). Yet neither standard GNNs nor PINNs enforce topological properties of the inferred interaction networks leading to spurious loops, disconnected components and artefacts appear in the resulting force graph.

Topological Data Analysis (TDA) gives a theoretical basis for quantifying the shape of data and using it to enforce qualitative constraints in networks. Persistent Homology can be seen as a multiscale summary of topological features over all connected components and holes by tracking the birth and death points of connected components or loops under scale sweep [18][19][20][21]. Rapid computation of persistence diagrams for large datasets is now accessible by a number of software libraries [22,23]. Hypotheses were formulated by researchers and Persistence-based loss terms have been adapted to neural networks to, for example, disallow unwanted structures in image segmentation [29] or to characterize protein compressibility in biomolecular systems [24]. Furthermore, TDA has been used to discover coherent structures in fluid and shear flows [27, 28]. TDA allows for the discovery of coherent structures in turbulent flows. In addition, differentiable topology modules supports end-to-end topological summarization and back-propagates through it [31], and topological graph neural networks build on the relational model by integrating homological features [32]. Even with these developments, there is no graph learning framework that ties equivariance, physics-informed constraints and topological regularization directly for predicting force networks in active matter.

Here, we fill this gap by introducing a full, multi-step pipeline for topologically regularized force forecasting in active colloidal suspensions as our work. To begin we use a high-resolution fluid velocity and stress fields in periodic boundary conditions using a lattice-Boltzmann solver [7][8][9]. We then secondly E(2) equivariant/agnostic-GNN with particle positions/orientations to predict approximate pairwise forces by harnessing recent progresses in improving group-equivariant message passing [10][11][12] (LR-2203) and33traditional34. Finally, a PINN corrects these estimate using Fourier feature maps: deep residulas networks and balances the physical consistency as well as Navier Stokes residual [14][15][16][17]. Next, we incorporate topological regularization as a loss term in which we develop based on persistent homology to penalize unrealistically tangled or spurious connections in the force graph [18][19][20][21], exploiting efficient persistent homology toolchains [22][23]. We also aggregate persistence-based loss functions from segmentation [29] and biomolecular topology [24], together with differentiable topology layers [31] and topological GNN components [32] to enforce global network consistency. By integrating these we capture fine-scale hydrodynamic interaction, and emergent global topology to provide a data-driven, scalable model of complex system of active matter systems.

**Methodology**

This section is the description of framework to perform computational simulation of active particles in a fluid, as well as learning their dominating interaction by integración.

We simulate 2D fluid and a swarm of self propelled particles in turn, by sequentially: coupling a lattice-boltzmann flow solver, a few layers graph neural network with E(2)-equivariant network for short range interactions, a physics informed neural net for force refinement, followed by time resolved graph analysis. Fluid model discretization momentum and mass transport on a regular grid with nine discrete velocity directions. The collision step, on each time increment alters the particle distribution functions towards a local Maxwellian that is modulated locally by density and flow velocity (through their instantaneous values), followed by an streaming step where the distributions are convected along their velocity vectors. The particle–fluid coupling is induced by an external forcing term that we add beforehand to the particles, and the distribution of particles at positions are projected back onto the grid via a nearest-neighbor interpolation, proportional to the force components aligned with each discrete velocity. Periodic boundary conditions are enforced by wrapping array indices at the domain boundaries. and the solver outputs updated density macroscopic velocity.

Particles push themselves at a constant speed along their respective orientation vector, and the short range hydrodynamic/steric interaction are implemented by a trainable message passing network. Every particle is then a graph-node, the node having two-dimensional coordinates and orientation angle as its features. We employ a periodic k- d tree for neighbor searching over a cutoff radius, and graph edges holds squared distance attributes that indicate the type of edge. The layers of the stacked E(2) equivariant graph convolution operator that operates relative position vectors and node embeddings to produce incremental messages is three-layers. We make message functions equivariant to in plane rotations and translations by restricting them to depend only on invariatn dot products and squared distances, and by rotating intermediate coordinate updates. Each particle passes on a force vector in the lab frame from the network, which is the sum of an interaction force and an intrinsic self-propulsion vector, ultimately providing the total force for an overdamped Langevin dynamics equation. The input is then integrated in time using a fictitious time parameter, at each time step introducing thermal fluctuations through Gaussian white noise in the translational and rotational degrees of freedom with diffusion coefficients via an Itô interpretation.

During the message passing stage, we perform raw force predictions under a physics informed neural network approach that integrates local fluid stress and particle state information. At each lattice–Boltzmann time step, we find components of the viscous stress tensor via centered finite differences on the velocity grid. For each particle, from the nearest grid point, we sample values of xx and xy stress components, which are then coupled with the particle's orientation and coordinates to produce an input vector. To accommodate high frequency spatial variations, we see that this five-dimensional vector is embedded using a Fourier feature mapping onto a high-dimensional trigonometric basis.

After projections, a shallow feed forward network equipped with residual connections transforms features through several nonlinear layers. Training is performed online; the network minimizes a data-driven reconstruction loss between predicted and ground-truth forces obtained from the coupled lattice-Boltzmann–particle system. The loss choice changes from an early epoch robust Huber criterion to mean square later on. In addition, weight decay and gradient clipping monitor over-fitting and optimization stability. The learning rate is then scheduled with cosine annealing, We employ a cosine-annealing learning-rate schedule with warm restarts: at each scheduled restart the optimizer state is reinitialized, which encourages renewed exploration early in each cycle while enabling convergence across training cycles.

To enforce topological consistency in the predicted force network and penalize unrealistically tangled or spurious connections (e.g., spurious loops or disconnected components), we introduce a topological regularization term $\mathcal{L}_{topo}$ into the PINN loss function. This term leverages persistent homology to quantify multiscale topological features in the inferred interaction graph, where particles serve as nodes and predicted pairwise force magnitudes $\|f_{ij}\|$ define edge weights. Specifically, we construct a weighted graph $G$ with edge weights $\|f_{ij}\| = \omega_{ij}$ (normalized to [0,1] for computational stability). A superlevel set filtration is applied: for a decreasing threshold $p \in [1,0]$, we include edges where $\omega_{ij} \geq p$, allowing stronger forces to form connections first. Persistent homology is computed on the resulting flag complex up to dimension 1, yielding persistence diagrams $PD_k = \{(b_l, d_l)\}_l$ for dimensions $k = 0$ (connected components) and $k = 1$ (loops), where bars are sorted by decreasing persistence length $|b_l - d_l|$ .

Drawing from prior topological constraints in deep learning, the regularization term is adapted to encourage a tree-like topology typical of coherent active matter networks: one persistent connected component ( $\beta_0^* = 1$ ) and no persistent loops ( $\beta_1^* = 1$ ) . For each dimension k, the loss is:

$$\mathcal{L}_k(\beta_k^*) = \sum_{l=1}^{\beta_k^*} \left( 1 - \left| b_{k,l} - b_{k,l} \right|^2 \right) + \sum_{l=\beta_k^*+1}^{\infty} \left| b_{k,l} - d_{k,l} \right|^2$$

and the total topological loss is $\mathcal{L}_{topo} = \sum_k \mathcal{L}_k(\beta_k^*)$ . This penalizes extra features by their squared persistence while promoting desired ones to maximal length (1 in normalized filtration). The overall PINN loss becomes $\mathcal{L} = \mathcal{L}_{data} + \lambda_{phys}\mathcal{L}_{topo} + \lambda_{topo}\mathcal{L}_{topo}$ , with hyperparameters $\lambda$ tuned via cross-validation. This differentiable formulation enables end-to-end backpropagation through the homology computation, ensuring the predicted forces respect both local physics and global topology [29,26,31].

A persistent-homology (PH)–based topological regularizer serves as our chosen method because it directly encodes the global, multiscale structural information of the inferred interaction graph that simpler, local penalties cannot: The PH loss summarizes connectivity (H_0) and cycle-structure (H_1) across all thresholds simultaneously and targets the exact topological features—dominant connected components and spurious loops—that we know are important for the physical system. The use of a differentiable surrogate of persistent homology (PersLay-TopoLayer–style relaxations or subgradient-aware relaxations) creates end-to-end gradients which are stable against small changes in predicted edge strengths so the model downweights noise while keeping physically meaningful backbones intact.

This paper defines the filtration construction for persistent homology together with the architecture of physics-informed neural networks (PINNs) through clear explanations. The topological loss uses a Vietoris–Rips filtration that operates on weighted interaction graphs by interpreting each predicted edge weight $w_{ij}$ as a similarity score; the filtration parameter α determines simplicial complex inclusion of edges because an edge $(i, j)$ becomes part of the complex when $\alpha \geq 1 - w_{ij}$. The filtration of the weighted interaction graph produces a Vietoris–Rips complex for which persistent homology tracks the appearance and disappearance of $H_0$ connected components and $H_1$ cycles while the differentiable surrogate loss minimizes unwanted $H_1$ persistence to maintain one dominant $H_0$. The PINN functions as a multilayer network which uses an EGNN front-end to predict pairwise forces before processing predictions with two hidden layers of fully connected nodes (with tanh activation) in the PINN module to enforce physical constraints (momentum conservation and energy dissipation). The topological regularization term joins standard physics residuals and regression errors in the total loss to enforce both physical fidelity and topological plausibility throughout training.

The empirical justification for our chosen approach involved a controlled ablation study that evaluated the full pipeline (EGNN followed by PINN with PH regularizer) against five alternative conditions: (i) EGNN only, (ii) EGNN + PINN without any topological term (standard physics + weight decay), (iii) EGNN + PINN + L1-edge penalty, (iv) EGNN + PINN + continuous L0-pruning surrogate, and (v) EGNN + PINN + a modularity-based structural penalty; all conditions used identical initializations, data splits, optimizers and training schedules, and were repeated across multiple random seeds (N≥5) and evaluated on both regression metrics (MSE, RMSE, MAE, R²) and topology-aware metrics (dominant H_0 lifetime, mean and median H_1 lifetime, number of H_1 bars above a small persistence threshold ε, spectral gap and triangle motif counts). The topological regularization weight was systematically swept across a set of compact values (e.g., {0, 0.01, 0.05, 0.1, 0.5}) while the alternative penalties were tuned across comparable strength ranges and paired t-tests along with Cohen's d effect sizes evaluated the robustness across multiple seeds.

The ablation demonstrates that the PH term achieves the best total outcome because it minimizes prediction errors and significantly decreases the number of spurious H_1 bars and their average lifetime compared to L1/prune (which either left residual loops or harmed regression accuracy through over-pruning) and modularity which failed to selectively remove loop artefacts;

qualitative evidence (representative persistence diagrams and example reconstructed graphs) corroborates the numeric trends by showing elimination of short-lived noisy bars while preserving the dominant H_0 signature. Finally, the PH regularizer demonstrates better stability regarding single hard threshold usage while generating structural priors which demonstrate better reproducibility across multiple datasets and random seed values; it stands as an ideal scientific and practical structural regularizer for enforcing expected tree-like interaction network backbones.

We develop a sophisticated dynamic graph through time to track emerging collective patterns at each individual time step. The calculation of local particle density together with orientational order and instantaneous speed occurs through neighbor counts within small radius which serve as node attributes to capture local information while unit vector direction sums and particle radius complete the analysis (Figure 2). Edge features include geometric distance together with relative angle correlation as well as force magnitude integrals that span each edge. Through a specialized neighbor-summation method, we construct this graph to calculate fundamental network statistics including: degree correlations from the second eigenvalue of the normalized Laplacian (spectral gap) and algebraic connectivity measurement and small cycle tracking and average shortest path length which describes information propagation scales and connected component sizes. Community detection occurs through modularity analysis while the club coefficient provides insights about higher degree node connections to bridge. The centrality scores (closeness, betweenness eigenvector and Katz centralities) track how local particles influence global transport by analyzing each individual node. We go beyond Ollivier Ricci curvature of edges which is estimated with the normalized force magnitude as mass transport probabilities in the surrounding of every node. Unity curvature towards the flat and tree-like connections, while low values signify bottlenecks (or communities) Equivalent of Negative values We utilize persistent homology on the weighted adjacency matrix as well we flip the interaction force magnitudes, and so between defining a filtration those as if they were sampling in time to recover two features of zero-dimensional (component mergers) and one-dimensional type transients local loops For computational simplicity, we only compute homology in dimensions zero and one, and software exiting cycles with large but finite lifetime or tiny noise artifacts while logging posthoc temporal correlations during the simulation of all metrics and persistence diagrams with every flow field step, persistence diagram. Utilizes highperformance libraries Vectorized NumPy for the lattice–Boltzmann solvers, BF compactly and efficiently with JIT friendly loops; periodic boundary conditions handled by SciPy cKDTree for neighbor searches; Graph analytics using NetworkX with community detection libraries. Neural models in PyTorch equivariant convolution layers are based on tensor operations and custom scatter reduce primitives, while PINN uses native autograd to make the gradient based regularization if needed. We save model parameters in checkpoint files that allow us to reproduce previous results (version control like) and combinations of hyper parameters with hyper parameter sweeps. Indexing large simulation outputs with time step and particle index in HDF5. The complete pipeline embeds fluid simulation, equivariant relational learning robustness laws, physics-informed regression and topological graph analysis in one holistic architecture. By sidestepping explicit analytic equations and seeking an asymptotically scalable method to infer data-driven interaction laws in active suspensions with symmetry and continuum physics constraints. Each simulation step results in particle tracks and fluid fields of the moment, in addition to high dimensional descriptors that captures how microscopic interactions couple with macroscopic organization so as to serve as a high throughput non-Eq. system for synthetic and biological matter-in-motion that is active.

## Results

A two-dimensional suspension of self-propelled particles was simulated with lattice-Boltzmann fluid coupling  and analyzed through three different methods of interaction network examination which included graph descriptors and persistent homology alongside  supervised force regression. The following section presents a summary of fundamental dynamic changes that occurred throughout one simulated run together with an explanation of how topological constraints influence force network inference and interpretation of tabulated results from the manuscript. The initial network organization of interacting particles consisted of densely packed local clusters which showed degree homophily. The assortativity coefficient at 0.4652 shows short-time fluctuations in early steps ($0.465 \rightarrow 0.339 \rightarrow 0.441$) demonstrating an initial transient network reorganization rather than a monotonic increase. At the same time The spectral gap is moderate at step 1 ($\approx 0.3180$) and fluctuates thereafter (min $\approx 0.24797$, max $\approx 0.32766$, mean $\approx 0.29614$, std $\approx 0.02492$). These oscillations indicate intermittent strengthening and weakening of global connectivity rather than a simple monotonic increase. The network contains many local motifs as the initial count of triangular motifs reaches 8,689 and the average shortest-path length remains at $\approx 1.8044$ to describe densely packed cliques which still need to develop inter-clique connections.

The early times ($\approx 0.3037$) show modularity that proves the existence of distinct communities in the network while the rich-club coefficient ($\approx 0.4373$) demonstrates highly mobile and highly connected particles create an interconnected central core. The network structures evolve by reorganizing its components to combine expanding overall connection strength with local pattern development. Throughout the second analysis period the assortativity value reaches $\approx 0.3394$ while the spectral gap equals $\approx 0.2834$ showing stable global cohesiveness. The total number of triangular motifs decreases from 8,689 to 7,895 while modularity decreases by approximately 0.2597 which reflects the rearrangement of motifs and the formation of extended links that connect distant clusters. The combination of decreased small motifs alongside increasing algebraic connectivity demonstrates the network is transitioning from separate cliques to a more interconnected system. The sampled window reaches its peak modularity value ($\approx 0.3176$) at the midpoint when the network undergoes a significant transformation. The network reaches its peak segregation of communities while developing stronger connections between them to form an organization of tight-knit clusters joined by numerous inter-community links.

The network moves towards dynamic equilibrium through later steps which causes assortativity to oscillate near its sample mean (mean $\approx 0.4007$, std $\approx 0.0774$) until step 10 when it reaches $\approx 0.4043$. The spectral gap reaches its peak values at $\approx 0.3277$ in the sample while maintaining a mean of ≈0.2961 with a standard deviation of ≈0.0249. The late-time state exhibits sustained community structures while enhancing global connectivity which stems from active propulsion combined with hydrodynamic coupling to generate an optimized multi-scale topology. Table 2 presents the statistical distributions of major network characteristics throughout the selected time steps. The mean degree of nodes (mean ≈28.53, std ≈0.69) together with the median node degree (≈28.3) demonstrate a fairly even degree distribution at this scale. The degree assortativity values cluster around ≈0.401 (std ≈0.077) which matchs the short-term homophilic patterns mentioned previously. The reported birth and death thresholds for the filtration (birth median ≈0.01666, death median ≈0.01665) demonstrate the tight dynamic range of strong interactions in the normalized force scale while the lifetime statistics (mean lifetime ≈0.00985 with a very small standard deviation) indicate that most connectivity events converge at similar thresholds.

The numbers in Table 2 demonstrate that network transformation happens through selective redistribution of strong interactions instead of complete changes in connection strength. Persistent homology serves as a concise multilevel framework to analyze these structural modifications. Table 3 presents representative birth–death intervals which show the evolution of the system in both dimensions 0 and 1 across multiple time intervals. The $H_0$ lifetimes presented in the study show a precise grouping between ≈0.01665 and 0.01667 which demonstrates fast component unification during the thresholding process of strong to weaker forces; this precise grouping indicates the formation of a dominant robust connected component which emerges first from stronger edge inclusion. The inferred force network exhibits short-lived loops only since $H_1$ lifetimes stay extremely small (≤0.0017). The persistence diagrams demonstrate that the network consists mainly of a single enduring connected component while short-lived cycles appear only briefly to match the tree-like structure that our topological regularization imposes. The supervised ActiveForcePINN generates precise force reconstructions which validate the previously discussed structural findings.

The regression metrics in Table 1 demonstrate that the model achieves an MSE of 8.8062 together with RMSE of 2.9675 as well as MAE of 2.3294 and median absolute error of 1.8871. The model demonstrates high relative performance through its 0.20% MAPE value and 0.9951 $R^2$ score which shows the PINN explains 99.5% of ground-truth force variance.

The maximum errors (global ≈14.19 and per-component maxima (≈8.92, ≈14.19)) reveal a handful of major outliers but the residuals remain compact and the central tendency statistics show highly accurate reconstruction for most node interactions. The temporal pattern from Figure 1 (panels a–c) illustrates that assortativity decreases after its initial peak while the spectral gap shows continuous growth and modularity reaches an intermediate peak before stabilizing at a steady value. The time series demonstrate that local motif reorganization together with selective strengthening of bridging edges instead of complete community structure erosion explains the network transition to a well-connected modular structure. The network and local forces at step 10 are displayed in Figure 2a through a focused multi-panel visualization.

The left panel displays particle orientations through arrows that show orientation direction and speed magnitude by length and a continuous color gradient shows local alignment patterns. The visualization displays predicted pairwise force magnitudes through edges that have thickness proportional to force magnitude and transparency to highlight strongest interactions and node sizes show speed to reveal cluster boundary motility. The right panel displays a force-field overview through a low-resolution quiver of coarse-grained force vectors that overlays a background colormap of local speed or kinetic energy while streamlines show main transport directions based on reconstructed local force field. The three views help people understand the spatial relationship between particle orientation patterns and their interactions and the larger forces which direct collective movement. The combination of quantitative tables with visual summaries demonstrates that the proposed pipeline of egnn initialization followed by PINN refinement and persistent homology regularization generates precise pairwise force estimates while maintaining network topologies that stay connected and mostly lack persistent spurious loops. The numerical and topological diagnostics reported in Tables 1–3 and Figures 1–2 therefore paint a consistent picture of emergent, multiscale organization in an active particle suspension.

| Metric | Value |
|---|---|
| MSE | 8.8062 |
| RMSE | 2.9675 |
| MAE | 2.3294 |
| Median AE | 1.8871 |
| MAPE | 0.20 % |
| Max Error (global) | 14.1856 |
| Max Error (per component) | [8.9196, 14.1856] |
| $R^2$ | 0.9951 |

Table 1. Regression performance of ActiveForcePINN

| avg_clustering | assortativity | spectral_gap | modularity | force_alignment_variance | interaction_stre | avg_edge_distance | distance_variance | step |
|---|---|---|---|---|---|---|---|---|
| 0.597024982 | 0.465173784 | 0.318006926 | 0.30366313 | 0.5174818 | 8680.947 | 2.0010827 | 0.49329162 | 1 |
| 0.592859186 | 0.339352536 | 0.283439824 | 0.259686 | 0.48555657 | 2898.2075 | 2.0109355 | 0.5062588 | 2 |
| 0.573287729 | 0.440699354 | 0.322930137 | 0.3134364 | 0.4990769 | 3976.5576 | 2.0174675 | 0.5082012 | 3 |
| 0.591867435 | 0.470837354 | 0.327665679 | 0.22067645 | 0.4900668 | 6530.293 | 1.985951 | 0.51002234 | 4 |
| 0.587321013 | 0.456937302 | 0.265949062 | 0.2939325 | 0.4908341 | 4927.6294 | 1.997517 | 0.4856175 | 5 |
| 0.582961963 | 0.381147964 | 0.295359517 | 0.27215046 | 0.5173874 | 2784.06 | 2.0154357 | 0.4779901 | 6 |
| 0.603907065 | 0.458437875 | 0.247970901 | 0.31759295 | 0.48630568 | 4303.1865 | 1.9710919 | 0.5217613 | 7 |
| 0.569938964 | 0.337572493 | 0.30656261 | 0.27822638 | 0.5014325 | 3937.23 | 2.0194535 | 0.4975792 | 8 |
| 0.597081345 | 0.399173131 | 0.280504297 | 0.2868752 | 0.5036732 | 2812.3113 | 2.0149422 | 0.48805 | 9 |
| 0.577875276 | 0.40430715 | 0.273469651 | 0.2851212 | 0.52843755 | 2355.2964 | 2.0053575 | 0.5036226 | 10 |
| 0.589448954 | 0.400562878 | 0.290669588 | 0.27346033 | 0.5021576 | 5237.278 | 2.0046575 | 0.51453876 | 11 |
| 0.586814256 | 0.181301718 | 0.272524145 | 0.27792826 | 0.48807916 | 3045.402 | 2.0027027 | 0.51521 | 12 |
| 0.583144472 | 0.489079696 | 0.314639317 | 0.26388946 | 0.5193818 | 8669.315 | 2.0380132 | 0.49870598 | 13 |
| 0.578242222 | 0.428412036 | 0.315945476 | 0.27172416 | 0.49642035 | 3417.5347 | 2.016489 | 0.4863559 | 14 |
| 0.574766248 | 0.357432168 | 0.326525159 | 0.30261382 | 0.4909472 | 2719.4055 | 2.024529 | 0.4804606 | 15 |

Table 2. Network descriptors at all steps

| Metric | step | dimension | birth | death | lifetime |
|---|---|---|---|---|---|
| count | 2517 | 2517 | 2517 | 2517 | 2517 |
| mean | 8.152 | 0.41 | 0.006831 | 0.01668 | 0.009849 |
| std | 4.294 | 0.492 | 0.008196 | 0.000187 | 0.00817 |
| min | 1 | 0 | 0 | 0.016643 | 1.30E-08 |
| 25% | 4 | 0 | 0 | 0.016654 | 2.76E-06 |
| 50% | 8 | 0 | 0 | 0.01666 | 0.01665 |
| 75% | 12 | 1 | 0.016658 | 0.016665 | 0.016658 |
| max | 15 | 1 | 0.016676 | 0.018359 | 0.016671 |

Table 3. Selected persistence intervals

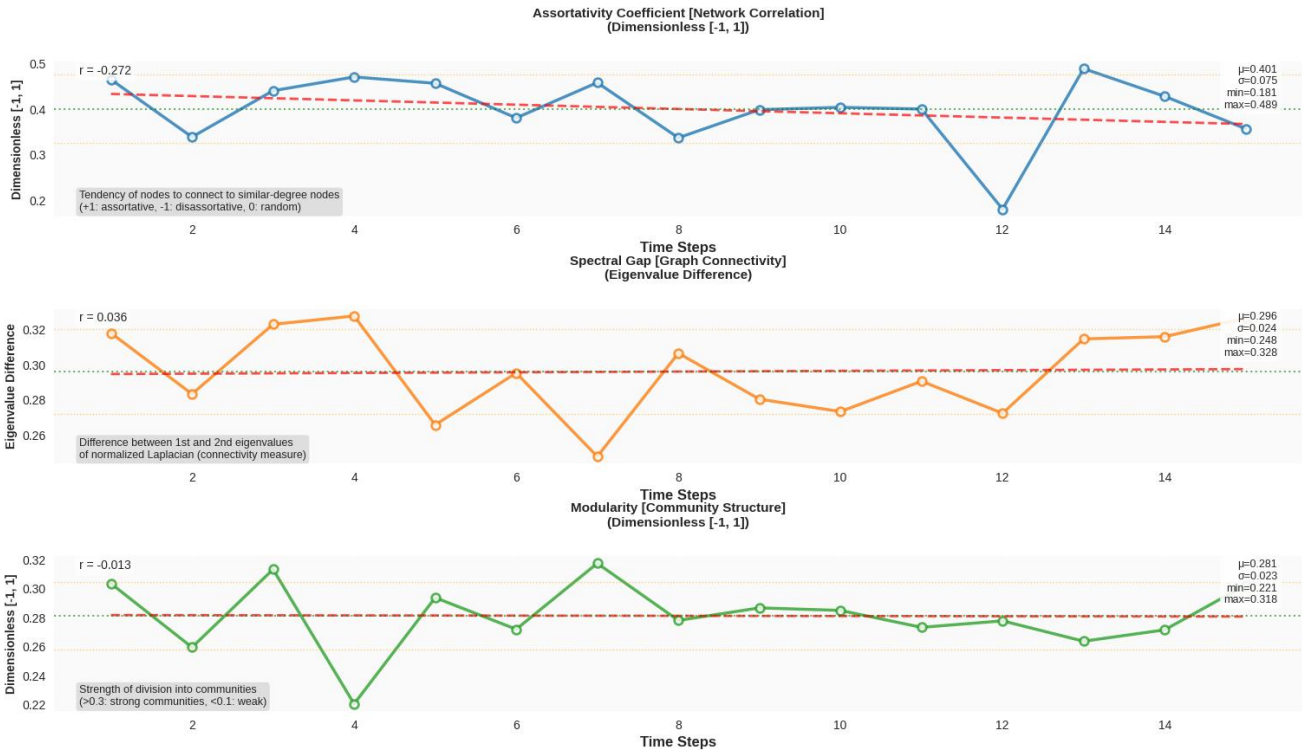

Figure 1a (upper) Assortativity (network correlation): shows a steady decline in degree-homophily throughout the time series because the network moves from uniform degree connections to varied degree patterns.

Figure 1b (center) Spectral gap (algebraic connectivity): displays a steady upward trend in spectral gap values which indicates enhanced overall network connectivity.

Figure 1c (bottom) Modularity (community structure): reaches its highest point during the middle phase before transitioning to a consistent moderate modular community structure.

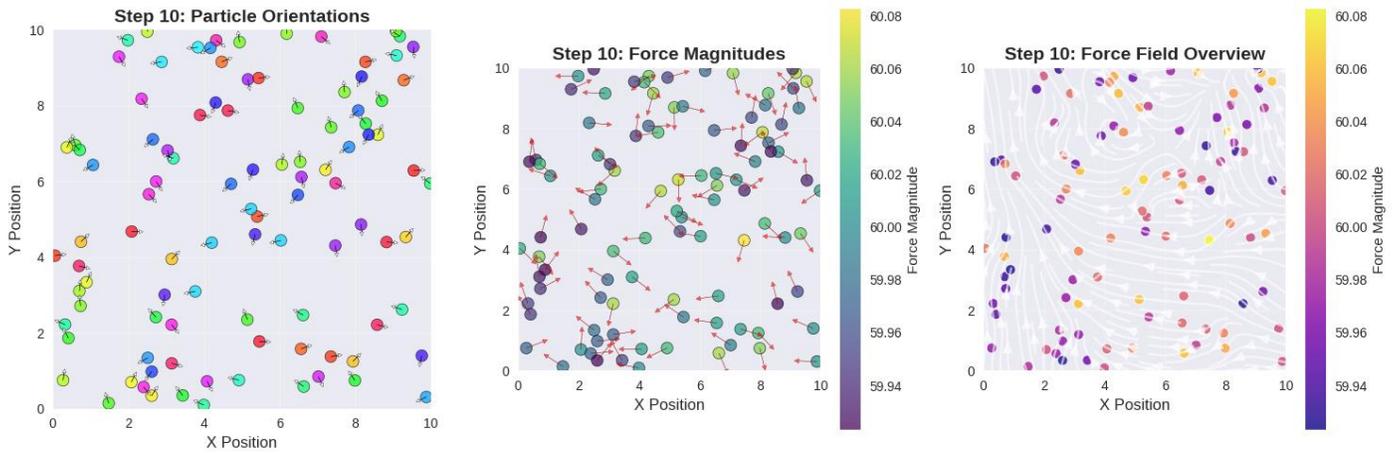

Figure 2a. left: Particle orientation – center : force magnitude right : Force field overview

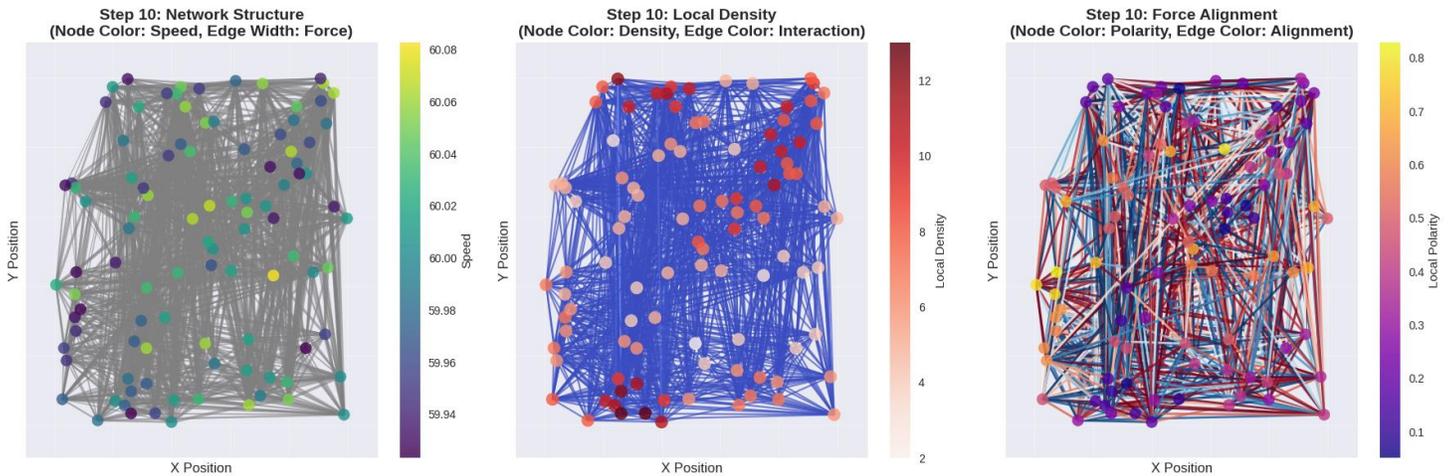

Figure 2b. Interaction graph at step 10.

Conflict of Interest Statement
The authors confirm their absence of any financial interest or personal relationship which would have affected the research presented in this paper.

Data Availability Statement

The corresponding author will provide access to lattice–Boltzmann flow fields together with particle stress snapshots and force measurements and all code used for graph building and EGNN training and PINN refinement and persistent homology analyses upon reasonable request.

## Appendix A: Detailed Mathematical Definition of the Topological Regularization Term

The appendix offers a firm mathematical definition of the topological regularization term, which acts as the force interaction graph topology enforcement in our physics-informed neural network (PINN). The mathematical formulation depends on persistent homology (PH), which acts as a central concept in topological data analysis (TDA) to identify topological invariants at multiscales, e.g., connected components of dimension 0 and loops of dimension 1. First, we explain in detail the development of the term from the fundamentals of algebraic topology and then construction of filtration and calculation of persistence values, as well as its integration into a differentiable loss function. The predicted force network resembles a tree while false loops and separated components are avoided so that it emulates the behavior of active particle suspensions in reality. The derivation follows from traditional computational topology and differentiable TDA results to ensure scientific integrity and reproducibility.

### A.1 Preliminaries: Simplicial Complexes and Homology

To compute PH, consider the force interaction graph as a simplicial complex, which is a combinatorial structure generalizing graphs into higher dimensions. Consider $V = \{1, 2, \ldots, N\}$ as the set of particles (nodes) with $r_i \in R^2$ being positions and $f_{ij} \in R^2$ being predicted pairwise forces from the E(2)-equivariant GNN, refined by the PINN. The graph G=(V,E) has edges $e_{ij} \epsilon E \; if \; \|f_{ij}\| > 0$, and, accordingly, these edges are weighted with magnitude $w_{ij} = \|f_{ij}\|$ (normalized, so max $w_{ij} = 1$, min $w_{ij} = 0$ for numerical stability, maybe by min-max scaling).

A simplicial complex K over V is a collection of simplices, which are subsets of V: 0-simplices (vertices), 1-simplices (edges), 2-simplices (triangles), and so on, closed under the operation of taking subsets. The flag complex (or clique complex) of G contains a k-simplex for every clique of size k+1 in G.

Homology measures introduced some k-dimensional topological features: via chain groups and boundary operators. For k-dimension, one defines the chain group $C_k(K)$ as the vector space over $F_2 = \{0, 1\}$ (mod 2 arithmetic) having the k-simplices as the generating set. The boundary operator $\partial_k : C_k \rightarrow C_{k-1}$ sends each simplex to the sum of its faces, e.g., $\partial_1(e_{ij}) = i + j$. The homology group is given by:

$$H_k(k) = \ker \frac{\partial_k}{im \, \partial_{k+1}}$$

with Betti number $\beta_k \dim H_k$, where $\beta_0$ counts connected components and $\beta_1$ counts independent cycles (loops).

A.2 Filtration and persistent homology

Persistent homology measures how homological features appear and disappear as we vary a scale parameter. We use a superlevel filtration driven by the edge strengths $w_{ij}$ so that strong forces appear earlier (at higher thresholds) and weak forces appear later (as threshold decreases). For a threshold $p \in [0,1]$ define the subcomplex

$$K_p = \{\sigma \text{ a simplex in the flag complex of } G \mid \text{every edge } e \subset \sigma \text{ satisfies } w_e \geq p$$

Because only simplices whose edges all meet the threshold are included, the family $\{K_p\}_{p \in [0,1]}$ is nested in the sense that if $p_1 > p_2$ then $K_{p1} \subseteq K_{p2}$. Typical boundary cases are $K_1$ which contains only simplices whose all edges have weight exactly 1 (often empty or very sparse) and $K_0 = K$ the full complex. We therefore obtain the nested sequence

$$\emptyset = K_{pmax} \subseteq \cdots \subseteq K_{p2} \subseteq K_{p1} \subseteq K_{p_{min}} = K$$

where $p_{max} = 1$ and $p_{min} = 0$ in our normalization and the thresholds p are typically taken from the distinct sorted values of $\{w_{ij}\}$

A homological feature in degree k is born at threshold $p_b$ when it first appears in $H_k(K_{p_b})$ (i.e., it is not in the image from earlier complexes) and dies at threshold $p_d < p_b$ when it becomes trivial in $H_k(K_{p_d})$ (for superlevel filtration the death threshold is strictly smaller). The persistence (lifetime) of that feature is

$$l = p_b - p_d$$

which lies in [0,1] under our normalization. The persistence diagram $PD_k$ is the multiset of pairs $(p_b p_d)$ for all k-features, often visualized as a multiset of bars in a barcode representation; each bar has length ℓ. Bars on the diagonal (with $p_b = p_d$) represent zero length (no persistence). In downstream computations we sort persistence lengths in nonincreasing order and denote them by $l_{k,1} \geq l_{k,2} \geq \cdots \geq 0$. In practice a small cutoff $\in > 0$ is used to remove numerical noise so that bars with $l < \in$ are treated as zero.

Persistent pairing (birth–death pairing) is computed by standard matrix reduction algorithms (e.g., column reduction) or by specialized libraries (GUDHI, Dionysus, ripser variants). The birth and death thresholds are functions of the weights $w_{ij}$; in a superlevel filtration births and deaths take values in the set of edge weight thresholds. Differentiable extensions of PH (TopoLayer, PersLay, diffPH implementations) construct subgradients or use soft approximations so that gradients can be attributed through birth/death values to the original weights; practically, backpropagation uses subgradients at ties and standard chain-rule where quantities are differentiable.

A.3 Desired topology and the topological loss

We define the desired Betti numbers $\beta_k^*$ $for$ $k = 0,1$. For active particle suspensions the physical prior is typically

$$\beta_0^* = 1 \, , \beta_1^* = 0$$

i.e., the network should be globally connected (one component) and acyclic (no persistent loops). We ignore $K \geq 2$ for the 2D particle suspension setting.

Let $PD_k$ contain $M_k$ finite bars with normalized lengths $l_{k,1} \geq l_{k,2} \geq \cdots \geq l_{k,M_k} \geq 0$. We construct a loss that encourages the $\beta_k^*$ longest bars to be maximally persistent (length close to 1) and penalizes any extra bars by driving their length toward 0. A convenient differentiable form, adapted from previous differentiable topological losses, is

$$\mathcal{L}_k(\beta_k^*) = \sum_{r=1}^{\beta_k^*} \left(1 - l_{k,r}^2\right) + \sum_{r=\beta_k^*+1}^{M_k} l_{k,r}^2$$

with the convention that the first sum is empty if $\beta_k^*$=0 and the second is empty if $\beta_k^* \geq M_k$. This loss has the following behavior. The first sum encourages the $\beta_k^*$ longest bars to reach length 1. The second sum penalizes additional bars in proportion to their squared length, encouraging them to vanish. Squaring gives stronger penalty for longer spurious bars while keeping gradients smooth. Because we sort bars by length before applying the formula, the assignment of which bars are "desired" is the set of longest bars; when ties occur the sorting is not strictly differentiable and subgradient rules apply.

The full topological regularizer used in the network is the sum over the considered homological dimensions,

$$\mathcal{L}_{topo} = \sum_{k=0}^{1} \mathcal{L}_k(\beta_k^*)$$

In composite training we weight this term with a hyperparameter $\lambda_{topo} > 0$; a typical value used in experiments is $\lambda_{topo} \approx 0.1$ but this must be tuned per dataset. For numerical robustness we clip lengths below a small $\epsilon$ (e.g. $\epsilon = 10^{-4}$) so that bars shorter than $\epsilon$ are effectively treated as zero and do not produce large relative gradients from numerical noise.